\documentstyle[a4,12pt]{article}
\begin{document}
\parskip=5pt plus 1pt minus 1pt

\begin{flushright}
{HEP-PH/9702297} \\
{\bf DPNU-96-62} \\
{\small December, 1996}
\end{flushright}

\vspace{0.2cm}

\begin{center}
{\Large\bf Towards Determining $\phi_1$ with $B\rightarrow D^{(*)}\bar{D}^{(*)}$} 
\end{center}

\vspace{0.3cm}

\begin{center}
{\bf A.I. Sanda} \footnote{Electronic address: sanda@eken.phys.nagoya-u.ac.jp} 
~ and ~
{\bf Zhi-zhong Xing} \footnote{Electronic address: xing@eken.phys.nagoya-u.ac.jp}
\end{center}
\begin{center}
{\it Department of Physics, Nagoya University, Chikusa-ku, Nagoya 464-01, Japan}
\end{center}

\vspace{2.5cm}

\begin{abstract}
We present an isospin analysis of the decay modes 
$B\rightarrow D\bar{D}$, $D^*\bar{D}$, $D\bar{D}^*$ and $D^*\bar{D}^*$,
which allows determination of the final-state interaction phases.
As these transitions have branching ratios of the order $10^{-4}$ or larger, 
they could be  useful in detecting the $CP$-violating phase $\phi_1 \equiv \arg 
(-V^*_{cb} V_{cd} V_{tb} V^*_{td} )$ in the first-round experiments of
a $B$-meson factory. The problem of penguin pollution may still be present.
Once the Kobayashi-Maskawa matrix elements
are known, it is possible to obtain the magnitudes and relative phases of
hadronic matrix elements for $B\rightarrow D^{(*)}\bar{D}^{(*)}$. This will in
turn lead to some information about the penguin pollution. 

\end{abstract}

\vspace{2cm}
\begin{center}
{ PACS number(s): ~ 13.25.+m, 11.30.Er, 12.15.Ff, 14.40Jz}
\end{center}

\newpage

\section{Introduction}

Within the standard electroweak model, three angles of the 
Kobayashi-Maskawa (KM) unitarity triangle
\begin{equation}
V^*_{ub}V_{ud} + V^*_{cb}V_{cd} + V^*_{tb}V_{td} \; = \; 0 \; ,
\end{equation}
denoted by $\phi_i$ ($i=1,2,3$), are determined from $CP$ asymmetries 
in neutral $B$-meson decays to hadronic $CP$ eigenstates \cite{Sanda,BigiSanda1,BigiSanda2}.
It is expected that $\phi_1 \equiv \arg (-V^*_{cb}V_{cd}V_{tb}V^*_{td} )$ 
can be unambiguously extracted from
the decay rate difference between $B^0_d\rightarrow \psi K_S$
and $\bar{B}^0_d\rightarrow \psi K_S$. While this decay mode
should be sufficient for determining $\phi_1$, the initial luminosity
of a $B$ factory may require us to search for some additional decay channels which could
help us establish the presence of $CP$ violation as quickly as
possible \cite{Review}. For this purpose, we shall investigate
\begin{equation}
B \; \longrightarrow \; D\bar{D} \; , \; D^*\bar{D} \; , \;
D\bar{D}^* \; , \; D^*\bar{D}^* \;
\end{equation}
in some detail.

In practical experiments the decay mode $B_d\rightarrow D^+D^-$ should
have fairly large branching ratio. Under SU(3) symmetry, one
can make the following rough estimation:
\begin{equation}
{\cal B} (B^0_d \rightarrow D^+D^-) \; \sim \; \sin^2\theta_{\rm C} ~
{\cal B} (B^0_d\rightarrow D^+_s D^-) \; \approx \; (3.4\pm 1.9) \times 10^{-4} \; ,
\end{equation}
where $\theta_{\rm C}$ is the Cabibbo angle, and ${\cal B}(B^0_d\rightarrow
D^+_sD^-) = (7\pm 4) \times 10^{-3}$ has been measured in experiments \cite{PDG}. 
Also, the penguin effect in $B_d\rightarrow D^+D^-$
is expected to be smaller than that in $B_d\rightarrow \pi^+\pi^-$.
Thus the $CP$ asymmetry between $B^0_d\rightarrow D^+D^-$ and
$\bar{B}^0_d\rightarrow D^+D^-$ may be dominated by angle $\phi_1$.
In contrast with $B_d\rightarrow D^+D^-$, 
the decay modes $B_d \rightarrow D^+ D^{*-}$, $D^{*+} D^-$ and 
$D^{*+}D^{*-}$ undergo the same weak interactions and have the comparable branching ratios,
although they are not the exact $CP$-even eigenstates \cite{France}.

Of course the measurement of $CP$ violation in $B_d\rightarrow 
D^+D^-$ can not only cross-check the extraction of $\phi_1$ from
$B_d\rightarrow \psi K_S$, but also shed some light on the penguin
effects and final-state interactions (FSIs) in nonleptonic $B$
decays to double charmed mesons. For this reason, it is worth studying
both $B_d\rightarrow D^+D^-$ and $B_d \rightarrow D^0\bar{D}^0$ 
in a model-independent approach. The similar treatment is applicable
to the processes $B_d\rightarrow D\bar{D}^*$, $D^*\bar{D}$, etc.

In this work we shall carry
out an isospin analysis of the processes $B\rightarrow D^{(*)}\bar{D}^{(*)}$,
to relate their weak and strong phases to the relevant observables.
It is found that the time-dependent measurements of $B_d\rightarrow
D^+D^-$ and $B_d\rightarrow D^0\bar{D}^0$ together with the 
time-independent measurements of $B^+_u\rightarrow D^+\bar{D}^0$
and $B^-_u\rightarrow D^-D^0$ allow one to extract a phase parameter $\phi'_1$,
which consists of both $\phi_1$ and the penguin-induced phase information.
Direct $CP$ asymmetries in $B_d\rightarrow D^+D^-$ and $D^0\bar{D}^0$
are time-independently
detectable on the $\Upsilon (4S)$ resonance. For numerical illustration,
we apply the effective weak Hamiltonian and factorization approximation
to $B^+_u\rightarrow D^{(*)+}\bar{D}^{(*)0}$ and $B^-_u\rightarrow D^{(*)-}D^{(*)0}$,
since each of them is only involved in a single isospin 
amplitude. We find that their branching ratios are all above $10^{-4}$ and the
relevant time-independent $CP$ asymmetries may reach the $3\%$ level. The
time-dependent $CP$ asymmetries in $B_d \rightarrow D^{(*)+}\bar{D}^{(*)-}$
and $D^{(*)0}\bar{D}^{(*)0}$ are expected to be of order 1.
We also emphasize that it is possible to obtain some information on the
magnitudes and relative phases of hadronic matrix elements in the isospin
approach, once the KM matrix elements have been known.

\section{Isospin analysis}

The effective weak Hamiltonians, responsible for 
$B^-_u\rightarrow D^-D^0$, $\bar{B}^0_d\rightarrow D^+D^-$,
$\bar{B}^0_d\rightarrow D^0\bar{D}^0$ and their $CP$-conjugate 
processes, have the isospin structures $|1/2, -1/2\rangle$
and $|1/2, +1/2\rangle$ respectively. The decay amplitudes of these 
transitions can be written in terms of the isospin amplitudes:
\begin{eqnarray}
A^{+-} & \equiv & \langle D^+D^- |{\cal H}_{\rm eff}| B^0_d\rangle \; = \;
\frac{1}{2} \left ( A_1 ~ + ~ A_0 \right ) \; , \nonumber \\
A^{00} & \equiv & \langle D^0\bar{D}^0 |{\cal H}_{\rm eff}| B^0_d\rangle \; = \;
\frac{1}{2} \left ( A_1  ~ - ~ A_0  \right ) \; , \nonumber \\
A^{+0} & \equiv & \langle D^+\bar{D}^0 |{\cal H}_{\rm eff}| B^+_u\rangle \; = \; A_1 \; ;
\end{eqnarray}
and
\begin{eqnarray}
\bar{A}^{+-} & \equiv & \langle D^+D^- |{\cal H}_{\rm eff}| \bar{B}^0_d\rangle \; = \;
\frac{1}{2} \left ( \bar{A}_1 ~ + ~ \bar{A}_0 \right ) \; , \nonumber \\
\bar{A}^{00} & \equiv & \langle D^0\bar{D}^0 |{\cal H}_{\rm eff}| \bar{B}^0_d\rangle \; = \;
\frac{1}{2} \left ( \bar{A}_1 ~ - ~ \bar{A}_0 \right ) \; , \nonumber \\
\bar{A}^{-0} & \equiv & \langle D^-D^0 |{\cal H}_{\rm eff}| B^-_u\rangle 
\; = \; \bar{A}_1 \; .
\end{eqnarray}
Here $A_1$ ($\bar{A}_1$) and $A_0$ ($\bar{A}_0$) are the isospin
amplitudes with $I=1$ and $I=0$, respectively. 
Clearly the isospin
relations (4) and (5) can be expressed as two triangles in the 
complex plane (see Fig. 1 for illustration):
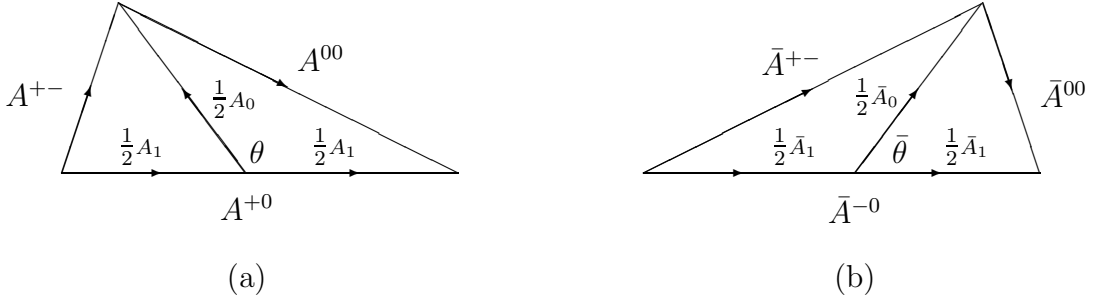
\begin{figure}
\begin{picture}(350,120)(20,250)
\put(80,300){\line(1,0){150}}
\put(113,300){\vector(1,0){5}}
\put(188,300){\vector(1,0){5}}
\put(150,287){\makebox(0,0){$A^{+0}$}}

\put(80,300){\line(1,3){21.5}}
\put(80,300){\vector(1,3){11}}
\put(70,330){\makebox(0,0){$A^{+-}$}}

\put(230,300){\line(-2,1){128}}
\put(101.5,364.5){\vector(2,-1){64}}
\put(178,343.5){\makebox(0,0){$A^{00}$}}

\put(101.5,364.5){\line(3,-4){48}}
\put(149.5,300){\vector(-3,4){24}}
\put(145,328){\makebox(0,0){$\frac{1}{2} \scriptstyle A_0$}}

\put(110,310){\makebox(0,0){$\frac{1}{2} \scriptstyle A_1$}}
\put(183,310){\makebox(0,0){$\frac{1}{2} \scriptstyle A_1$}}
\put(154,308){\makebox(0,0){$\theta$}}

\put(150,260){\makebox(0,0){(a)}}


\put(300,300){\line(1,0){150}}
\put(333,300){\vector(1,0){5}}
\put(408,300){\vector(1,0){5}}
\put(380,286){\makebox(0,0){$\bar{A}^{-0}$}}

\put(300,300){\line(2,1){128}}
\put(300,300){\vector(2,1){64}}
\put(356,342){\makebox(0,0){$\bar{A}^{+-}$}}

\put(450,300){\line(-1,3){21.5}}
\put(428.5,364.5){\vector(1,-3){11}}
\put(459,330){\makebox(0,0){$\bar{A}^{00}$}}

\put(428,364){\line(-3,-4){48}}
\put(380,300){\vector(3,4){24}}
\put(388,329){\makebox(0,0){$\frac{1}{2} \scriptstyle \bar{A}_0$}}

\put(357,310){\makebox(0,0){$\frac{1}{2} \scriptstyle \bar{A}_1$}}
\put(422,310){\makebox(0,0){$\frac{1}{2} \scriptstyle \bar{A}_1$}}
\put(397,309){\makebox(0,0){$\bar{\theta}$}}

\put(380,260){\makebox(0,0){(b)}}
\end{picture}
\caption{The isospin triangles of $B\rightarrow D\bar{D}$ in the complex plane.}
\end{figure}
\begin{eqnarray}
A^{+-} ~ + ~ A^{00} & = & A^{+0} \; , \nonumber \\
\bar{A}^{+-} ~ + ~ \bar{A}^{00} & = & \bar{A}^{-0} \; .
\end{eqnarray}
One is able to determine the relative size and phase difference of 
isospin amplitudes $A_1$ ($\bar{A}_1$) and $A_0$ ($\bar{A}_0$) from the
above triangular relations. Denoting
\begin{equation}
\frac{A_0}{A_1} \; \equiv \; z e^{{\rm i}\theta}\; , ~~~~~~~~
\frac{\bar{A}_0}{\bar{A}_1} \; \equiv \; \bar{z} 
e^{{\rm i} \bar{\theta}} \; ,
\end{equation}
then we obtain
\begin{eqnarray}
z & = & \sqrt{\frac{2 \displaystyle \left ( |A^{+-}|^2 + |A^{00}|^2 \right )}
{|A^{+0}|^2} ~ - ~ 1} \; , \nonumber \\
\theta & = & \arccos \left ( \frac{|A^{+-}|^2 - |A^{00}|^2}
{z ~ |A^{+0}|^2} \right ) \; ; 
\end{eqnarray}
and
\begin{eqnarray}
\bar{z} & = & \sqrt{\frac{2 \left ( |\bar{A}^{+-}|^2 + |\bar{A}^{00}|^2 \right )}
{|\bar{A}^{-0}|^2} ~ - ~ 1} \; , \nonumber \\
\bar{\theta} & = & \arccos \left ( \frac{|\bar{A}^{+-}|^2 - |\bar{A}^{00}|^2}
{\bar{z} ~ |\bar{A}^{-0}|^2} \right ) \; .
\end{eqnarray}
If $z=1$ and $\theta =0$, for example, we find that $|A^{00}|=0$, i.e.,
the decay mode $B^0\rightarrow D^0\bar{D}^0$ is forbidden.

Note that $\theta$ ($\bar{\theta}$) is in general a mixture of the weak and
strong phase shifts, since both $A_0$ ($\bar{A}_0$) and $A_1$ ($\bar{A}_1$)
may contain the tree-level and penguin contributions. This point can be 
seen more clearly if one writes the isospin amplitudes $A_I$ and $\bar{A}_I$
($I=1, 0$) with the help of the low-energy effective $\Delta B =\pm 1$ 
Hamiltonians. For example, $A_I$ can be given as 
\begin{equation}
A_I \; = \; \langle (D\bar{D})_I | {\cal H}_{\rm eff} (\Delta B = +1) | B\rangle 
\; = \; \frac{G_F}{\sqrt{2}} \sum_{q=u,c} \left [ \left (V_{qb}^* V_{qd} \right )
S^q_I \right ] 
\end{equation}
with
\begin{equation}
S^q_I \; = \; c_1 \langle (D\bar{D})_I |Q^q_1 |B\rangle ~ + ~ 
c_2 \langle (D\bar{D})_I |Q^q_2 |B\rangle  ~ + ~
\sum^{10}_{i=3} \left [ c_i \langle (D\bar{D})_I |Q_i |B\rangle \right ] \; ,
\end{equation}
where Wilson coefficients $c_i$ and four-quark operators $Q_i$ at the scale
$\mu = O(m_b)$ have been well defined in Ref. \cite{Buras}. 
The expression of $\bar{A}_I$ is straightforwardly obtainable from Eq. (10)
through the replacement $(V_{qb}^*V_{qd}) \rightarrow (V_{qb}V_{qd}^*)$.
The tree- and penguin-type hadronic matrix elements in $S^u_I$
are expected to consist of different strong phases, and these phases should 
be different from those in $S^c_I$. This implies that the overall phases of 
$A_1$ ($\bar{A}_1$) and $A_0$ ($\bar{A}_0$) are non-linear combinations of 
the same weak phases and the different strong phases, therefore $\theta$ 
($\bar{\theta}$) is neither purely weak nor purely strong.

Finally it is worth mentioning that the same isospin relations hold for the decay
modes $B\rightarrow D\bar{D}^*$ and $B\rightarrow D^*\bar{D}$. Of course,
the isospin parameters $z$ ($\bar{z}$) and $\theta$ ($\bar{\theta}$) in
$B\rightarrow D\bar{D}$, $D\bar{D}^*$ and $D^*\bar{D}$ may be different 
from one another due to their different FSIs. As for $B\rightarrow D^*\bar{D}^*$,
the same isospin relations hold separately for the decay amplitudes
with helicity $\lambda=-1$, $0$, or $+1$.

\section{Time-independent measurements}

The quantities $|A^{+0}|$ and $|\bar{A}^{-0}|$ are obtainable from
the time-independent measurements of 
decay rates of $B^+_u\rightarrow D^+\bar{D}^0$ and
$B^-_u\rightarrow D^-D^0$. A determination of
$|A^{+-}|$ ($|A^{00}|$) and $|\bar{A}^{+-}|$ ($|\bar{A}^{00}|$) is possible
through the time-integrated measurements of $B^0_d$ vs $\bar{B}^0_d\rightarrow 
D^+D^-$ ($D^0\bar{D}^0$) 
on the $\Upsilon (4S)$ resonance, where the produced two $B_d$ mesons are
in a coherent state (with odd charge-conjugation parity) until one of them 
decays. In practice, one can use 
the semileptonic transition of one $B_d$ meson to tag the flavor 
of the other meson decaying to $D^+D^-$ or $D^0\bar{D}^0$. The probability for
observing such a joint decay event reads \cite{BigiSanda2,Xing2}:
\begin{equation}
{\cal R}(l^{\pm}X^{\mp}; D^+D^-) \; \propto \; |A_l|^2 \left ( \frac{|A^{+-}|^2
+ |\bar{A}^{+-}|^2}{2} ~ \mp ~ \frac{1}{1+x^2_d} \cdot \frac{|A^{+-}|^2
- |\bar{A}^{+-}|^2}{2} \right ) \; , 
\end{equation}
or
\begin{equation}
{\cal R}(l^{\pm}X^{\mp}; D^0\bar{D}^0) \; \propto \; |A_l|^2 \left ( \frac{|A^{00}|^2
+ |\bar{A}^{00}|^2}{2} ~ \mp ~ \frac{1}{1+x^2_d} \cdot \frac{|A^{00}|^2
- |\bar{A}^{00}|^2}{2} \right ) \; , 
\end{equation}
where $|A_l| \equiv |\langle l^+ X^- |{\cal H}_{\rm eff}| B^0_d \rangle | = |\langle l^-X^+
|{\cal H}_{\rm eff}| \bar{B}^0_d \rangle |$ under $CPT$ symmetry, and $x_d = \Delta m/\Gamma
\approx 0.73$ is a measure of $B^0_d -\bar{B}^0_d$ mixing \cite{PDG}. By now the semileptonic
$B_d$ transitions such as $B^0_d\rightarrow D^{(*)-}l^+\nu^{~}_l$ and 
$\bar{B}^0_d\rightarrow D^{(*)+}l^-\bar{\nu}^{~}_l$ have
been well reconstructed \cite{PDG}, i.e., $|A_l|$ has been detected independent of the above
joint decay modes. Once ${\cal R}(l^{\pm}X^{\mp}; D^+D^-)$ and ${\cal R}(l^{\pm}X^{\mp}; 
D^0\bar{D}^0)$ are measured, we shall be able to determine the quantities 
$|A^{+-}|$ ($|A^{00}|$) and $|\bar{A}^{+-}|$ ($|\bar{A}^{00}|$). 

The time-independent measurements mentioned above allow one to construct the
isospin triangles in Fig. 1. Consequently the isospin parameters $z$ ($\bar{z}$)
and $\theta$ ($\bar{\theta}$) are extractable in the absence of any 
time-dependent measurement. If the branching ratios of $B^0_d\rightarrow 
D^0\bar{D}^0$ and $\bar{B}^0_d\rightarrow D^0\bar{D}^0$ are too small to be
observable, then large cancellation between the isospin amplitudes $A_1$
($\bar{A}_1$) and $A_0$ ($\bar{A}_0$) must take place. In the case that
$B^0_d\rightarrow D^+D^-$ and $B^+_u\rightarrow D^+\bar{D}^0$ have been
measured earlier than $B^0_d\rightarrow D^0\bar{D}^0$, a lower bound on the
rate of the latter decay mode is model-independently achievable from the isospin
relations obtained above. Since $\cos\theta \leq 1$, we get from Eq. (8) that
\begin{equation}
{\cal B} (B^0_d\rightarrow D^0\bar{D}^0) \; \geq \; \left [ \sqrt{\frac{
{\cal B} (B^0_d\rightarrow D^+D^-)}{{\cal B} (B^+_u\rightarrow D^+\bar{D}^0)}} ~ - ~ 1
\right ]^2 {\cal B} (B^+_u\rightarrow D^+\bar{D}^0) \; ,
\end{equation}
where tiny isospin-violating effects induced by the mass difference 
$m^{~}_{D^0} - m^{~}_{D^-}$ and the life time difference $\tau^{~}_{B_d}
- \tau^{~}_{B_u}$ have been neglected. This bound should be useful to 
set a limit for the results of ${\cal B} (B^0_d\rightarrow D^0\bar{D}^0)$ obtained
from specific models of hadronic matrix elements.
Following the same way, one can find the lower bounds for the branching ratios of
$B^0_d\rightarrow D^{*0}\bar{D}^0$, $D^0\bar{D}^{*0}$ and $D^{*0}\bar{D}^{*0}$.

The nonvanishing asymmetry between the decay rates of $B^+_u\rightarrow D^{+}\bar{D}^{0}$
and $B^-_u\rightarrow D^{-}D^{0}$ signifies the existence of direct $CP$ violation.
By use of the isospin amplitudes in Eqs. (10) and (11), we obtain the following
$CP$ asymmetry:
\begin{equation}
{\cal A}_{\pm 0} \; \equiv \; \frac{{\cal R}(B^+_u\rightarrow D^{+}\bar{D}^{0})  - 
{\cal R}(B^-_u\rightarrow D^{-}D^{0})}{{\cal R}(B^+_u\rightarrow D^{+}\bar{D}^{0})  + 
{\cal R}(B^-_u\rightarrow D^{-}D^{0})} \; = \; 2 \sin\phi_3 
~ \frac{{\rm Im} (S^u_1 S^{c*}_1)}{N_{11}} \; ,
\end{equation}
where $\phi_3 \equiv \arg (-V^*_{ub}V_{ud}V_{cb}V^*_{cd})$ is an angle of the KM unitarity triangle,
and $N_{11}$ can be read from 
\begin{equation}
N_{ij} \; \equiv \; \kappa ~ {\rm Re} (S^u_i S^{u*}_j )
~ + ~ \kappa^{-1} ~ {\rm Re} (S^c_i S^{c*}_j ) ~ - ~ \cos\phi_3 ~ 
{\rm Re} (S^u_i S^{c*}_j + S^u_j S^{c*}_i ) \; 
\end{equation}
with $\kappa \equiv |V_{ub}V_{ud}|/|V_{cb}V_{cd}|$.
For the processes $B_d\rightarrow D^{+}D^{-}$ and $D^{0}\bar{D}^{0}$, pure signals of direct 
$CP$ asymmetries may manifest themselves {\it on the $\Upsilon (4S)$ resonance}:
\begin{eqnarray}
{\cal A}_{+-} & \equiv & \frac{{\cal R}(l^-X^+; D^{+}D^{-}) ~ - ~ 
{\cal R}(l^+X^-; D^{+}D^{-})}{{\cal R}(l^-X^+; D^{+}D^{-}) ~ + ~ 
{\cal R}(l^+X^-; D^{+}D^{-})} \; , \nonumber \\
& = & \frac{2 \sin\phi_3}{1+x^2_d} \cdot \frac{{\rm Im} (S^u_1 S^{c*}_1 
+ S^u_0 S^{c*}_0 + S^u_1 S^{c*}_0 + S^u_0 S^{c*}_1 )}{N_{11} + N_{00} + N_{10} + N_{01}} \; ,
\end{eqnarray}
and
\begin{eqnarray}
{\cal A}_{00} & \equiv & \frac{{\cal R}(l^-X^+; D^{0}\bar{D}^{0}) ~ - ~ 
{\cal R}(l^+X^-; D^{0}\bar{D}^{0})}{{\cal R}(l^-X^+; D^{0}\bar{D}^{0}) ~ + ~
{\cal R}(l^+X^-; D^{0}\bar{D}^{0})} \; \nonumber \\
& = & \frac{2 \sin\phi_3}{1+x^2_d} \cdot \frac{{\rm Im} (S^u_1 S^{c*}_1 
+ S^u_0 S^{c*}_0 - S^u_1 S^{c*}_0 - S^u_0 S^{c*}_1 )}{N_{11} + N_{00} - N_{10} - N_{01}} \; .
\end{eqnarray}
If the decay modes $B^0_d \rightarrow D^0\bar{D}^0$ and $\bar{B}^0_d\rightarrow
D^0\bar{D}^0$ are forbidden due to the absence of final-state rescattering 
(i.e., $\theta \approx 0$ and $z\approx 1$, or $S^q_1 \approx 
S^q_0$), then measuring the $CP$ asymmetry ${\cal A}_{00}$ is practically impossible. 
In this case, we arrive at an interesting relation between the asymmetries
${\cal A}_{\pm 0}$ and ${\cal A}_{+-}$:
\begin{equation}
{\cal A}_{\pm 0} \; \approx \; \left (1 + x^2_d \right ) {\cal A}_{+-} \; 
\approx \; 1.5 {\cal A}_{+-} \; .
\end{equation}
The validity of this relation is testable in the forthcoming
experiments at a $B$-meson factory.

It is worthwhile at this point to give a brief comparison between the isospin
language and the intuitive quark-diagram description for $B\rightarrow D^{(*)}
\bar{D}^{(*)}$. Both the isospin amplitudes $A_1$ and $A_0$ are dominated by 
the spectator (external $W$-emission) quark graph with the KM factor $V^*_{cb}V_{cd}$, 
but they also receive some small contributions from the loop-induced penguin
and annihilation-type tree quark diagrams. Hence the branching ratios of 
$B^0_d\rightarrow D^{(*)+}D^{(*)-}$ and $B^+_u\rightarrow D^{(*)+}\bar{D}^{(*)0}$
may be of the same order. In the assumption of no final-state rescattering or
channel mixing, $B^0_d\rightarrow D^{(*)0}\bar{D}^{(*)0}$ takes palce only through
the annihilation-type quark graphs, which are expected to have significant formfactor 
suppression in the factorization approximation. This argument is compatible with
the isospin analysis, since the cancellation between $A_1$ and $A_0$ in $A^{00}$ 
implies that the dominant spectator diagram does not contribute to 
$B^0_d\rightarrow D^{(*)0}\bar{D}^{(*)0}$. However, one should keep in mind that
FSI effects are possible to significantly enhance the decay rate of $B^0_d\rightarrow 
D^{(*)0}\bar{D}^{(*)0}$ to the level comparable with that of
$B^0_d\rightarrow D^{(*)+}D^{(*)-}$ or $B^+_u\rightarrow D^{(*)+}\bar{D}^{(*)0}$,
making the naive quark-diagram language in failure. 

\section{Time-dependent measurements}

To probe the $CP$ asymmetry induced by the interplay of direct decay
and $B^0_d-\bar{B}^0_d$ mixing in $B_d\rightarrow D\bar{D}$, the time-dependent
measurements are necessary on the $\Upsilon (4S)$ resonance at asymmetric
$B$ factories. In such an experimental scenario, the joint decay rates can be given as
follows \cite{Xing2}:
\begin{eqnarray}
{\cal R}(l^{\pm}X^{\mp}, D^+D^-; t) & \propto & |A_l|^2 e^{-\Gamma |t|} \left [
\frac{|A^{+-}|^2 + |\bar{A}^{+-}|^2}{2} ~ \mp ~ \frac{|A^{+-}|^2 - |\bar{A}
^{+-}|^2}{2} \cos (x_d \Gamma t) \right . \nonumber \\
&  & \left . \pm ~ |A^{+-}|^2 ~ {\rm Im} \left ( \frac{q}{p}
\frac{\bar{A}^{+-}}{A^{+-}} \right ) \sin (x_d \Gamma t) \right ] \; 
\end{eqnarray}
and
\begin{eqnarray}
{\cal R}(l^{\pm}X^{\mp}, D^0\bar{D}^0; t) & \propto & |A_l|^2 e^{-\Gamma |t|} \left [
\frac{|A^{00}|^2 + |\bar{A}^{00}|^2}{2} ~ \mp ~ \frac{|A^{00}|^2 - |\bar{A}
^{00}|^2}{2} \cos (x_d \Gamma t) \right . \nonumber \\
&  & \left . \pm ~ |A^{00}|^2 ~ {\rm Im} \left ( \frac{q}{p}
\frac{\bar{A}^{00}}{A^{00}} \right ) \sin (x_d \Gamma t) \right ] \; , 
\end{eqnarray}
where $t$ is the proper time difference between the semileptonic and nonleptonic
decays
\footnote{Note that the proper time sum of the semileptonic and nonleptonic 
decays has been integrated out, since it will not be measured at any $B$-meson factory.},
and 
\begin{equation}
\frac{q}{p} \; \equiv \; \left | \frac{q}{p} \right | \exp (-{\rm i} 2\phi_0)
\; \approx \; \exp (-{\rm i} 2\phi_0) \;
\end{equation}
stands for the phase information from 
$B^0_d-\bar{B}^0_d$ mixing \cite{BigiSanda2}. For simplicity, we denote the phase difference
between $A_1$ and $\bar{A}_1$ as 
\begin{equation}
\varphi \; \equiv \; \frac{1}{2} \arg \left (\frac{\bar{A}_1}{A_1} \right )
\; = \; \frac{1}{2} \arg \left [ \frac{V_{cb}V^*_{cd}}{V^*_{cb}V_{cd}} \cdot
\frac{S^c_1 - \kappa S^u_1 \exp (-{\rm i} \phi_3)}{S^c_1 - \kappa S^u_1
\exp(+{\rm i} \phi_3)} \right ] \; .
\end{equation}
In terms of the isospin parameters, coefficients of the $\sin (x_d \Gamma t)$ term
in Eqs. (20) and (21) are given by
\begin{eqnarray}
{\rm Im} \left ( \frac{q}{p} \frac{\bar{A}^{+-}}{A^{+-}} \right ) & = &
\frac{|A^{+0} \bar{A}^{-0}|}{4 |A^{+-}|^2} \left [ - \sin \left (2\phi'_1 \right ) ~ - ~
z \sin \left (\theta + 2\phi'_1 \right ) \right . \nonumber \\
&  & \left . + ~ \bar{z} \sin \left (\bar{\theta} - 2\phi'_1 \right ) ~ + ~ z \bar{z} 
\sin \left (\bar{\theta} - \theta - 2\phi'_1 \right ) \right ] \; 
\end{eqnarray}
and
\begin{eqnarray}
{\rm Im} \left ( \frac{q}{p} \frac{\bar{A}^{00}}{A^{00}} \right ) & = &
\frac{|A^{+0} \bar{A}^{-0}|}{4 |A^{00}|^2} \left [ - \sin \left (2\phi'_1 \right ) ~ + ~
z \sin \left (\theta + 2\phi'_1 \right ) \right . \nonumber \\
&  & \left . - ~ \bar{z} \sin \left (\bar{\theta} - 2\phi'_1 \right ) ~ + ~ z \bar{z} 
\sin \left (\bar{\theta} - \theta - 2\phi'_1 \right ) \right ] \; ,
\end{eqnarray}
where $\phi'_1 \equiv \phi_0 - \varphi$. All the quantities on the right-hand side of
Eq. (24) or (25), except $\phi'_1$, can be determined through the time-independent
measurements of $B\rightarrow D\bar{D}$ on the $\Upsilon (4S)$ resonance. Thus 
measuring the $CP$-violating observable on the left-hand side of Eq. (24) or (25)
will allow a model-independent extraction of $\phi'_1$.

Let us make two remarks about the results obtained above:

(1) Within the standard model, $\phi_0 \approx \phi_1$ holds to an excellent degree
of accuracy. If the tree-level quark transition
$\bar{b}\rightarrow (c\bar{c})\bar{d}$ dominates the decay amplitude of 
$B^+_u\rightarrow D^+\bar{D}^0$ (i.e., $|S^c_1| \gg |S^u_1|$ in Eq. (23)),
then we get $\varphi\approx \arg (V_{cb}V^*_{cd}) \approx 0$ as a pure weak
phase. In this case, the magnitude of $\phi_1$ is extractable from the time-dependent
measurement of $B_d\rightarrow D^+D^-$ or $B_d\rightarrow D^0\bar{D}^0$ \cite{Sanda,BigiSanda1}.
Note that a model-dependent estimation in the standard model gives $\varphi \sim -3^{\circ}$
(see the appendix). 
It is worth pointing out that the $B^0_d-\bar{B}^0_d$ mixing phase $\phi_0$ can be
reliably determined from the $CP$ asymmetry in $B^0_d$ vs $\bar{B}^0_d\rightarrow \psi K_S$ 
either within or beyond the standard model
\footnote{This point relies on the condition that no significant penguin effect
exists in $B_d\rightarrow \psi K_S$, and it is testable through the time-independent
measurement of the decay rate difference between $B^0_d\rightarrow \psi K_S$
and $\bar{B}^0_d\rightarrow \psi K_S$ on the $\Upsilon (4S)$ resonance.}.
Thus a comparison of $\phi_0$ (extracted from $B_d\rightarrow \psi K_S$) with
$\phi'_1$ (extracted from $B_d\rightarrow D^+D^-$ or $B_d\rightarrow D^0\bar{D}^0$) 
will constrain $\varphi$, which may reflect the penguin-induced phase information 
in $B\rightarrow D\bar{D}$.

(2) It is interesting to note that we can, in principle, obtain phases
and magnitudes of the hadronic matrix elements $S^q_I$.
On the experimental side, $|A_0|$ ($|\bar{A}_0|$), $|A_1|$ ($|\bar{A}_1|$) 
and $\theta$ ($\bar{\theta}$) 
can be determined from the time-independent measurements; and
$\phi_1'$ can be extracted from the time-dependent measurements.
If the KM phases are known, the hadronic matrix elements $S^u_0$,
$S^u_1$, $S^c_0$ and $S^c_1$ represent seven unknown parameters 
as the overall phases of them are physically irrelevant.
Thus the magnitudes and relative phases of these quantities
should be determinable from experimental measurements of the relevant branching 
ratios and $CP$ asymmetries.

A special but interesting case is $z = \bar{z} =1$. It can be obtained if 
the decay modes $B\rightarrow D\bar{D}$ occur dominantly through the tree-level
subprocess $b\rightarrow (c\bar{c}) d$ or $\bar{b}\rightarrow (c\bar{c}) \bar{d}$.
In this case, $A_0$ ($\bar{A}_0$) and $A_1$ ($\bar{A}_1$) have a common KM factor;
thus $\theta$ ($\bar{\theta}$) is a 
pure strong phase shift. This will lead, for arbitrary values of $\theta$ and 
$\bar{\theta}$, to the results
\begin{eqnarray}
| A^{+-}| & = & | A^{+0}| \cos\frac{\theta}{2} \; , ~~~~~~~~
| A^{00}| \; = \; |A^{+0}| \sin\frac{\theta}{2} \; , \nonumber \\
| \bar{A}^{+-}| & = & | \bar{A}^{-0}| \cos\frac{\bar{\theta}}{2} \; , ~~~~~~~~
| \bar{A}^{00}| \; = \; |\bar{A}^{-0}| \sin\frac{\bar{\theta}}{2} \; .
\end{eqnarray}
As a straightforward consequence, one gets
\begin{eqnarray}
|A^{+-}|^2 ~ + ~ |A^{00}|^2 & = & |A^{+0}|^2 \; , \nonumber \\
|\bar{A}^{+-}|^2 ~ + ~ |\bar{A}^{00}|^2 & = & |\bar{A}^{-0}|^2 \; , 
\end{eqnarray}
i.e., the two isospin triangles in Fig. 1 become right-angled triangles.
If $\theta = \bar{\theta}$ is further assumed, we obtain
\begin{eqnarray}
{\rm Im} \left ( \frac{q}{p} \frac{\bar{A}^{+-}}{A^{+-}} \right ) & = &
- \frac{|A^{+0} \bar{A}^{-0}|}{|A^{+-}|^2} \sin \left (2\phi'_1 \right ) 
\cos^2 \frac{\theta}{2} \; , \nonumber \\
{\rm Im} \left ( \frac{q}{p} \frac{\bar{A}^{00}}{A^{00}} \right ) & = &
- \frac{|A^{+0} \bar{A}^{-0}|}{|A^{00}|^2} \sin \left (2\phi'_1 \right ) 
\sin^2 \frac{\theta}{2} \; .
\end{eqnarray}
One can see that these two $CP$-violating quantities have the quasi-seesaw 
dependence on the isospin phase shift $\theta$. The magnitude of $\sin (2\phi'_1)$
turns out to be
\begin{equation}
\sin(2\phi'_1) \; = \; - \frac{1}{|A^{+0}\bar{A}^{-0}|} \left [
|A^{+-}|^2 {\rm Im}\left (\frac{q}{p} \frac{\bar{A}^{+-}}{A^{+-}} \right )
~ + ~ |A^{00}|^2 {\rm Im} \left (\frac{q}{p} \frac{\bar{A}^{00}}{A^{00}} \right )
\right ] \; ,
\end{equation}
apparently independent of $\theta$.
 
\section{Summary and conclusion}

We have presented an isospin analysis of the weak decays $B\rightarrow D^{(*)}\bar{D}^{(*)}$.
The main results can be summarized as follows:

(a) The time-independent measurements 
of these transitions on the $\Upsilon (4S)$ resonance allow one to extract the isospin 
quantities and probe the direct $CP$ asymmetries in them. It is possible to extract
a phase parameter, which consists of the phase information from both $B^0_d-\bar{B}^0_d$
mixing and penguin diagrams, from the time-dependent measurements of $B_d\rightarrow
D^+D^-$ and $D^0\bar{D}^0$. A comparison of this phase with that extracted from 
$B_d\rightarrow \psi K_S$ will be interesting, since their difference signifies 
the penguin-induced phase information (no matter whether new physics is present or not).

(b) Once the KM matrix elements have been determined, the relevant hadronic matrix elements
(including their phase information) can be determined, through the isospin analysis,
from some measurements of the decay rates and $CP$ asymmetries.

In the appendix, we have made use of the effective weak Hamiltonian and naive factorization
approximation to estimate the branching ratios of $B^+_u\rightarrow D^{(*)+}\bar{D}^{(*)0}$
and $B^-_u\rightarrow D^{(*)-}D^{(*)0}$ as well as their $CP$ asymmetries. 
It is remarkable that all these decay modes can well be detected in the first-round
experiments of a $B$-meson factory. In particular,
only about $10^8$ $B^{\pm}_u$ events are expected to need for the exploration of
direct $CP$-violating signals in them (at the $3\%$ level).

We conclude that a careful experimental study of the decay modes 
$B\rightarrow D^{(*)}\bar{D}^{(*)}$ at the forthcoming $B$ factories, 
before the measurements of $B\rightarrow \pi\pi$ 
and other charmless $B$ decays become available, 
will be able to cross-check the extraction of the weak angle $\phi_1$ from
$B_d\rightarrow \psi K_S$, to shed some light on the penguin and FSI effects in
B decays to double charmed mesons, and to probe direct $CP$ violation in both 
charged and neutral $B$-meson systems.

\vspace{0.5cm}
\begin{flushleft}
{\Large\bf Aknowledgments}
\end{flushleft}

One of the authors (A.I.S.) likes to acknowledge the Daiko Foundation for
a partial support to his research. Z.Z.X. is grateful to Yoshihito Iwasaki
for helpful discussions about the detectability of $B\rightarrow D\bar{D}$
at a $B$ factory, and to the Japan Society for the Promotion of
Science for its financial support. 


\vspace{0.6cm}
\begin{flushleft}
{\Large\bf Appendix}
\end{flushleft}

Here we calculate the branching ratios of $B^+_u \rightarrow
D^{(*)+}\bar{D}^{(*)0}$ and $B^-_u\rightarrow D^{(*)-}D^{(*)0}$ as well as their 
$CP$ asymmetries numerically, in order to give one a feeling of ballpark numbers 
to be expected within the standard model.
It is suitable to apply the effective weak Hamiltonian and factorization
approximation to these decay modes, because each of them only has a single
isospin amplitude. In contrast, $B_d\rightarrow D^{(*)+}D^{(*)-}$ or $B_d\rightarrow 
D^{(*)0}\bar{D}^{(*)0}$ is involved in two different isospin amplitudes; thus a direct
application of the factorization approximation to such transitions may be 
problematic unless the FSI effects are negligibly small.

In estimating the branching ratios of $B^+_u\rightarrow D^+\bar{D}^0$,
$D^{*+}\bar{D}^0$, $D^+\bar{D}^{*0}$ and $D^{*+}\bar{D}^{*0}$, 
it is instructive to neglect small contributions from the hadronic matrix elements 
$\langle D^+\bar{D}^0|Q^u_{1,2}|B^+_u\rangle$ (annihilation) and 
$\langle D^+\bar{D}^0|Q_{3-10}|B^+_u\rangle$ (penguin).
These transitions have the weak interaction similar to that in 
$B^+_u\rightarrow D^+_s\bar{D}^0$, $D^{*+}_s\bar{D}^0$, $D^+_s\bar{D}^{*0}$
and $D^{*+}_s\bar{D}^{*0}$, whose decay rates have already been measured in
experiments \cite{PDG}. Then a comparison between the above two sets of
decay modes, with the help of the factorization approximation, leads
straightforwardly to the following leading-order results:
$$
{\cal B} (B^+_u\rightarrow D^+\bar{D}^0) \; \approx \; \frac{f^2_{D^+}}{f^2_{D^+_s}}
~ \sin^2\theta_{\rm C} ~ {\cal B} (B^+_u\rightarrow D^+_s \bar{D}^0) \; , 
$$
$$
{\cal B} (B^+_u\rightarrow D^+\bar{D}^{*0}) \; \approx \; \frac{f^2_{D^+}}{f^2_{D^+_s}}
~ \sin^2\theta_{\rm C} ~ {\cal B} (B^+_u\rightarrow D^+_s \bar{D}^{*0}) \; , 
$$
$$
{\cal B} (B^+_u\rightarrow D^{*+}\bar{D}^0) \; \approx \; \frac{g^2_{D^{*+}}}{g^2_{D^{*+}_s}}
~ \sin^2\theta_{\rm C} ~ {\cal B} (B^+_u\rightarrow D^{*+}_s \bar{D}^0) \; , 
$$
$$
{\cal B} (B^+_u\rightarrow D^{*+}\bar{D}^{*0}) \; \approx \; \frac{g^2_{D^{*+}}}{g^2_{D^{*+}_s}}
~ \sin^2\theta_{\rm C} ~ {\cal B} (B^+_u\rightarrow D^{*+}_s \bar{D}^{*0}) \; ,
\eqno({\rm A1}) 
$$
where $\theta_{\rm C}$ is the Cabibbo angle, $f_X$ and $g^{~}_{X^*}$ ($X=D^+$ or $D^+_s$) 
are the decay constants. Since our present knowledge of $f_{D^+}$, $f_{D^+_s}$,
$g^{~}_{D^{*+}}$ and $g^{~}_{D^{*+}_s}$ is quite poor \cite{PDG}, we take
$f_{D^+} \approx f_{D^+_s} \approx 0.8$ and $g^{~}_{D^{*+}} \approx g^{~}_{D^{*+}_s} 
\approx 0.8$ for simplicity and illustration \cite{Rosner}. 
Choosing the central values of ${\cal B} (B^+_u\rightarrow D^+_s \bar{D}^0)$,
etc \cite{PDG}, we approximately obtain ${\cal B}(B^+_u\rightarrow D^+\bar{D}^0)
\approx 5.3 \times 10^{-4}$, ${\cal B}(B^+_u\rightarrow D^+\bar{D}^{*0}) \approx
3.1\times 10^{-4}$, ${\cal B}(B^+_u\rightarrow D^{*+}\bar{D}^0) \approx 3.7 \times
10^{-4}$ and ${\cal B}(B^+_u\rightarrow D^{*+}\bar{D}^{*0}) \approx 7.1 \times 10^{-4}$.
From this rough estimation one can see that the above decay modes are definitely
detectable in the first-round experiments of a $B$-meson factory.

To roughly estimate the $CP$ asymmetry between $B^+_u\rightarrow D^+\bar{D}^0$ and
$B^-_u\rightarrow D^-D^0$, we take the time-like penguin contribution
into account \cite{Penguin}. The annihilation and space-like penguin effects are expected to be
negligible if we insist on the significant formfactor suppression associated with them  
\footnote{However, one should keep in mind that such an argument may not be
on a solid ground and has to be examined after some theoretical (experimental)
progress is made in deeper understanding of the dynamics of nonleptonic $B$ decays.}.
Then the overall decay amplitudes can be calculated, by use of the QCD-improved
effective weak Hamiltonian and factorization approximation, in a renormalization-scheme
independent way \cite{Fleischer,BSW}. Instead of repeating the technical details of 
such a treatment, here we only write out the resultant expressions of $S^u_1$ and
$S^c_1$ in the assumptions made above:
$$
S^u_1 \; \propto \; \left ( \frac{\bar{c}_3}{3} + \bar{c}_4 + \frac{\bar{c}_9}{3}
+ \bar{c}_{10} \right ) ~ + ~ \left ( \frac{\bar{c}_5}{3} + \bar{c}_6 + \frac{\bar{c}_7}{3}
+ \bar{c}_8 \right ) \xi_c 
$$
$$
~~~~~~~~~~ + ~ \frac{1 + \xi_c}{9\pi} \left [ \bar{c}_2 \alpha_s + \left ( \bar{c}_1 +
\frac{\bar{c}_2}{3} \right ) \alpha_e \right ] \left [ \frac{10}{9} +
F_u (k^2) \right ] \; , 
$$
$$
S^c_1 \; \propto \; \left ( \frac{\bar{c}_1}{3} + \bar{c}_2 + 
\frac{\bar{c}_3}{3} + \bar{c}_4 + \frac{\bar{c}_9}{3}
+ \bar{c}_{10} \right ) ~ + ~ \left ( \frac{\bar{c}_5}{3} + \bar{c}_6 + \frac{\bar{c}_7}{3}
+ \bar{c}_8 \right ) \xi_c  
$$
$$
+ ~ \frac{1 + \xi_c}{9\pi} \left [ \bar{c}_2 \alpha_s + \left ( \bar{c}_1 +
\frac{\bar{c}_2}{3} \right ) \alpha_e \right ] \left [ \frac{10}{9} +
F_c (k^2) \right ] \; ,
\eqno({\rm A2})
$$
where the common hadronic matrix element $\langle D^+|(\bar{c}d)_{V-A}|0\rangle
\langle \bar{D}^0|(\bar{b}c)_{V-A}|B^+_u\rangle$ has been singled out from 
$S^u_1$ and $S^c_1$. 
In Eq. (A2), $\alpha_s$ and $\alpha_e$ are the strong and electroweak coupling constants
respectively; $\bar{c}_i$ stands for the renormalization-scheme independent Wilson coefficient;
$\xi_c = 2m^2_{D^+}/[m_c (m_b - m_c)]$ arises from the transformation of
$(V-A)(V+A)$ currents into $(V-A)(V-A)$ ones for $Q_{5-8}$; and $F_q(k^2)$ denotes
the penguin loop-integral function with momentum transfer $k$ at the scale
$\mu = O(m_b)$:
$$
F_q (k^2) \; = \; 4 \int^1_0 {\rm d} x ~ x (1-x) \ln \left [ \frac{m^2_q -
k^2 x (1-x)}{m^2_b} \right ] \; .
\eqno({\rm A3})
$$
The absorptive part of $F_q(k^2)$ emerges for $k^2 \geq 4 m^2_q$, leading to
the possibility of direct $CP$ violation \cite{Penguin}.

One can calculate $S^u_1$ and $S^c_1$ for the decay modes $B^+_u\rightarrow
D^{*+}\bar{D}^0$, $D^+\bar{D}^{*0}$ and $D^{*+}\bar{D}^{*0}$ using the same
factorization approximation. If the polarizations of final-state vector mesons are summed
over, we arrive at the same formulas as Eq. (A2) with $\xi_c =0$ for
$D^{*+}\bar{D}^0$, $\xi_c = -2m^2_{D^+}/[m_c (m_b + m_c)]$ for $D^+\bar{D}^{*0}$,
and $\xi_c =0$ for $D^{*+}\bar{D}^{*0}$. Of course, such results depend upon 
the assumptions made above and cannot be taken too seriously.

With the help of Eqs. (A2) and (A3), one is able to evaluate the $CP$ asymmetry
${\cal A}_{\pm 0}$ defined in Eq. (15) and the phase parameter $\varphi$ given
in Eq. (23). For illustration, we typically choose $m_u=5$ MeV, $m_c=1.35$
GeV, $m_b=5$ GeV and $m_t=174$ GeV. The Wolfenstein parameters are taken to be
$\lambda=0.22$, $A=0.81$, $\rho=0.05$ and $\eta=0.36$. We adopt values of the 
Wilson coefficients $\bar{c}_i$ obtained in Ref. \cite{He}. 
The unknown penguin momentum transfer $k^2$ is treated as a free parameter changing
from $0.01m^2_b$ to $m^2_b$. 
A few points can be drawn from the explicit numerical calculations:

(a) The QCD (gluonic) penguin plays the dominant role in the overall penguin
amplitude, while the electroweak penguin effect is negligibly small.
At $k^2=4m^2_c\approx 0.3 m^2_b$ both ${\cal A}_{\pm 0}$ and $\tan (2\varphi)$ 
undergo a remarkable change in magnitude.

(b) The $CP$ asymmetries ${\cal A}_{\pm 0}$ between $B^+_u\rightarrow D^{(*)+}\bar{D}^{(*)0}$ and 
$B^-_u\rightarrow D^{(*)-} D^{(*)0}$ have the same sign and are of the order $3\%$. 
The relative change of each asymmetry due to the uncertain penguin momentum transfer 
$k^2$ is less than $15\%$. 

(c) With the inputs listed above, the phase parameter $\varphi$ is estimated to 
be around $-3^{\circ}$. Considering the large uncertainties associated with the 
inputs and the approach itself, we believe that a significant deviation of $\varphi$
from zero (e.g., $\varphi \sim -10^{\circ}$) cannot be excluded even within the
standard model.

(d) Observation of the above $CP$-violating signals to three standard 
deviations needs about $10^8$ $B^{\pm}_u$ events, if the
composite detection efficiency is at the $10\%$ level. Such measurements 
are possible in the first-round experiments at the forthcoming $B$ factories.

In the case that the decay channels $B_d \rightarrow D^{(*)0}\bar{D}^{(*)0}$ 
are significantly suppressed,
we expect that the direct $CP$ asymmetries in $B_d\rightarrow D^{(*)+}D^{(*)-}$
are comparable in magnitude with those in $B^+_u\rightarrow D^{(*)+}\bar{D}^{(*)0}$
vs $B^-_u\rightarrow D^{(*)-}D^{(*)0}$ (see Eq. (19) for illustration).
Nevertheless, much more $B^0_d\bar{B}^0_d$ events are needed to detect the former
on the $\Upsilon (4S)$ resonance due to the cost for flavor tagging. It is likely to
measure direct $CP$-violating signals in $B_d\rightarrow D^{(*)+}D^{(*)-}$
in the second-round experiments of a $B$-meson factory.

\end{document}